\documentstyle[12pt]{article}

\begin{document}
\begin{titlepage}
\title{MODEL STUDY OF HOT AND DENSE BARYONIC MATTER}
\author{Abhijit Bhattacharyya and Sanjay K. Ghosh \\
Department of Physics, Bose Institute, \\ 93/1, A. P. C. Road, 
Calcutta 700 009, INDIA \\email: phys@boseinst.ernet.in}
\maketitle
\begin{abstract}
The properties of baryonic matter have been investigated at finite density 
and temperature using different models. The variation of baryon masses and 
fractional number densities with baryon density and temperature obtained 
from different models have been compared. The quark hadron phase transition 
have been studied using Chiral Colour Dieletric (CCD) model in the quark 
sector. No phase transition has been seen for the different variants of the 
Zimanyi-Moszkowski model. However, a phase transition is observed for the 
linear and non-linear Walecka model.
\end{abstract}
\end{titlepage}
\newpage

\section{Introduction}
The study of hot and dense hadronic matter is a very old sphere of 
activity \cite{1a}. It is quite well known now that QCD is the fundamental
theory of strong interaction where the fundamental particles are quarks and 
gluons. In principle, one should be able to study the low energy hadronic  
phenomena using QCD. But due to the nonperturbative nature of the problem,
it becomes difficult to use QCD to describe hadronic matter. This leads 
people to calculate hadronic matter properties from effective models. 

The quark structure of hadrons lead people to believe that under extreme 
conditions, {\it i.e.} at 
very high temperature and/or density, the quarks, which are confined inside 
the hadrons, may have larger spatial extension \cite{1j,1k} compared to the 
typical size of a hadron which is about $1 fm$. So hadron to quark matter 
phase transition may occur inside neutron stars because of the very high 
density. On the other hand, a soup of quarks and 
gluons may be produced in heavy ion collisions. Such a matter was 
termed as Quark-Gluon Plasma (QGP) \cite{1k}. Independent of the fact 
whether QGP will be produced, it can be said that a hot and dense matter 
will be produced under such extreme conditions. That can be either QGP or 
hot and dense hadronic matter. Furthermore, even if QGP is produced it 
will undergo a phase transition to hadronic matter. These phenomena make 
the study of hadronic matter an interesting area of research. 

In this paper we have studied the properties of baryonic matter at high 
temperature and density from the five different hadronic models. These are 
the Walecka model (LW), the Zimanyi-Moszkowski (ZM) models and the non-linear 
Walecka (NW) model. The Walecka model \cite{1h} contains nucleons, 
scalar $\sigma$-meson and vector $\omega$-meson and it reproduces the 
features of strong interaction {\it i.e.} short range repulsive and long 
range attractive forces. Extensive studies have been carried out, after 
the invention of this model, towards the calculation of hadronic matter 
properties from this model \cite{1a,3a}. This revealed certain 
inconsistencies in the model. It was found that 
this model yields the incompressibility $K = 524 MeV$ compared to the 
accepted value $K = 210 \pm 30 MeV$. Furthermore the model gives a 
very low nucleon mass at the nuclear matter density. As a remedy of these 
problems other models of hadronic matter were proposed. In the NW \cite{1af} 
model, in addition to the interactions present in the usual Walecka 
model, cubic and quartic self interactions of the $\sigma$-meson are also
included. 

The ZM model \cite{1ah} differs from the usual Walecka model in the form 
of coupling between the scalar meson and the nucleon, which is a 
derivative coupling in the new model. As a result of this derivative 
coupling the model reproduces some of the experimental results 
nicely \cite{1ah,3d}. It yields the incompressibility $K = 224.49 
MeV$, which is much closer to the accepted value compared to the 
Walecka model, and a nucleon effective mass $M_N^* = 797.64 MeV$ at the 
nuclear matter saturation density $(\rho = 0.15 fm^{-3})$. But, the price 
one has to pay is that the model looses its renormalisability due to the 
derivative coupling. This is acceptable as far as the 
discussion of the effective models goes, since the description of hadronic 
matter need only be valid upto the temperatures $T \le 200MeV$ provided the 
deconfinement phase transition to QGP is a reality. 

In the present work the LW and ZM models have been extended to include 
hyperons. Some efforts have already been made to include hyperon degrees of 
freedom in the nuclear equation of state \cite{1af,1ag,1ai,1aj} but, most 
of these studies were restricted to zero temperature scenario. Section 2 
is devoted to the study of different models of baryonic 
matter. In section 3 the qurak model, which is chosen to be the Chiral 
Colour Dielectric (CCD) model has been discussed. Section 4 contains 
the study of the temperature and density dependence of baryon masses and 
number densities. The hadron-quark phase transition is discussed in section 
5 and finally the results and conclusions are given in section 6. 

\section{Hadronic Models} 
We have five different hadronic models to describe the properties of baryonic 
matter. These models are the Linear Walecka model, the three variants of the 
Zimanyi-Moszkowski model and the Non-linear Walecka model. The different 
variants of the ZM model is denoted by ZM, ZM2 and ZM3 (according 
to the notations followed in ref.\cite{3d,3e}) models respectively. The ZM 
models discussed in refs. \cite{3d,3e} contain neutrons and protons only and 
they have been used to describe symmetric nuclear matter. Here the ZM 
models as well as the linear Walecka model are extended to include 
hyperons, $\rho$-meson and leptons (electrons and muons). The 
$\rho$-meson and leptons have been included to describe the asymmetric 
baryonic matter in $\beta$-equilibrium. 

The Lagrangian density for the NW model \cite{1af} is written as,
\begin{eqnarray}
{\cal L}_{NW} = \sum_{i} \bar \psi_{i} (i\gamma^{\mu}\partial_{\mu}
- m_i +g_{\sigma i}
\sigma+ g_{\omega i} \omega_{\mu} \gamma^{\mu}- 
g_{\rho i} \rho^{a}_{\mu} \gamma^{\mu} T_{a} ) \psi_{i} 
-{1 \over {4}} \omega ^{\mu \nu} \omega_{\mu \nu} \nonumber \\ 
+{1 \over {2}} m^{2}_{\omega} \omega_{\mu} \omega^{\mu}  
+ {1 \over {2}} ( \partial_{\mu} \sigma \partial^{\mu} \sigma-
m^{2}_{\sigma} \sigma^{2}) 
- {1 \over {4}} \rho^{a}_{\mu \nu} \rho^{\mu \nu}_{a} +
{1 \over {2}} m_{\rho}^{2} \rho^{a}_{\mu} \rho^{\mu}_{a} \nonumber\\
 - {1 \over {3}} bm_{N} (g_{\sigma N} \sigma)^{3} -
{1 \over {4}} c( g_{\sigma N} \sigma )^4 +
\sum_{l}\bar\psi_{l}(i\gamma^{\mu}\partial_{\mu}-m_l)\psi_{l}
\label{lnw}
\end{eqnarray}

The Lagrangian densities fo LW model and ZM models can be written in a 
unified form as given below \cite{1h,1ah,3d,3e}: 
\begin{eqnarray} 
{\cal L}_{R} &=& \sum_i {\bar \psi}_i (i \gamma_\mu \partial^\mu  - M_i 
+ m^{* \beta} g_{\sigma i} \sigma - 
m^{* \alpha} g_{\omega i} \gamma_\mu \omega^\mu + 
m^{* \alpha} g_{\rho i} \gamma_\mu \rho^\mu_a T_a) \psi_i \nonumber\\ 
&+& m^{* \alpha} ( {1 \over 4} \omega_{\mu \nu} \omega^{\mu \nu} - 
{1 \over 2} m_\omega^2 \omega_\mu \omega^\mu ) 
+ {1 \over 2}(\partial_\mu \sigma \partial^\mu \sigma - m^2 
\sigma^2) \nonumber\\
&-& m^{* \alpha} ({1 \over {4}} \rho^{a}_{\mu \nu} \rho^{\mu \nu}_{a} - 
{1 \over {2}} m_{\rho}^{2} \rho^{a}_{\mu} \rho^{\mu}_{a}) 
+ \sum_l {\bar \psi}_l (i \gamma_\mu \partial^\mu  - M_l) \psi_l 
\label{lrs}
\end{eqnarray}
with $\alpha = 0, \beta = 0$ for the Walecka model, $\alpha = 0, 
\beta = 1$ for the ZM model, $\alpha = 1, \beta = 1$ for the ZM2 model, 
and $\alpha = 2, \beta = 1$ for the ZM3 model.   

In the Lagrangians (\ref{lnw}) and (\ref{lrs}) mentioned above, $\psi_i$ 
is the baryon field which can be neutron (n), proton (p), lambda ($\Lambda$) 
or sigma ($\Sigma^-$) baryons, $\omega$, $\rho$ 
and $\sigma$ are the corresponding meson fields, $g_{\sigma i}$, 
$g_{\omega i}$ and $g_{\rho i}$ are the coupling strengths of $\sigma$, 
$\omega$ and $\rho$ mesons, respectively, with the corresponding 
baryons. The electrons and muons are denoted by $\psi_l$. The Non-linear 
Walecka Lagrangian includes cubic and quartic self-interactions of the 
$\sigma$ field with couplings $b$ and $c$ respectively \cite{1af}. The 
quantity $m^*$ is given by 
\begin{equation}
m^* = (1 + {{g_{\sigma n} \sigma_0} \over M_n})^{-1}
\end{equation}
where $g_{\sigma n}$ is the coupling of scalar meson with nucleon, 
$M_n$ is the nucleon mass and $\sigma_0$ is the mean field value of 
$sigma$ field. 

In the Mean Field Approximation (MFA) the meson fields are replaced 
by their ground state expectation values. Then the equation of motion for 
the baryon field can be written as 
\begin{equation}
\left[i \gamma .\partial - M_i^* - g_{\omega i} \gamma_0 \omega^0 + T_3 
\gamma_0 \rho^{30} \right] = 0
\end{equation}
where $T_3$ is the z-component of the isospin and $M_i^*$ is the effective 
mass of the baryon which is  
\begin{equation}
M_i^* = M_i - g_{\sigma i} m^{* \beta} \sigma_0
\end{equation}
for the linear Walecka and ZM models. For the NW model $M_i^*$ can be 
written as,
\begin{equation}
M_i^* = M_i - g_{\sigma i} \sigma_0
\end{equation}

The mean field values of the meson fields are given by 
\begin{eqnarray}
\omega_0 &=& \sum_i {{g_{\omega i}} \over {m_\omega^2}} 
\langle {\bar \psi}_i \gamma_0 \psi_i \rangle \nonumber\\
\rho_0^3 &=& \sum_i {{g_\rho} \over {m_\rho^2}} 
\langle {\bar \psi}_i I^3 \gamma_0 \psi_i \rangle 
\end{eqnarray}
for all the models. The value of $\sigma_0$ for the 
Lagrangian (\ref{lrs}) is 
\begin{equation}
\sigma_0 = \sum_i {{g_{\sigma i}} \over {m^2_\sigma}} 
\langle {\bar \psi}_i \psi_i \rangle + {\alpha \over 2} \left [ {m_\omega 
\over m_\sigma} \right]^2 {\alpha \over M} m^{*(\alpha + 1)} \omega_0^2
\end{equation}
and for the non-linear Walecka model (\ref{lnw}) is
\begin{equation}
\sigma_0 = \sum_i {{g_{\sigma i}} \over {m^2_\sigma}} 
\langle {\bar \psi}_i \psi_i \rangle - b m_N g_{\sigma N}^3 \sigma_0^2 
- c g_{\sigma N}^4 \sigma_0^3
\end{equation}

The Lagrangian (\ref{lrs}) has six parameters. These are $g_{\sigma n} /
m_{\sigma}$, $g_{\sigma \Lambda} / m_{\sigma}$, $g_{\rho n} / m_\rho$ and 
$g_{\rho \Lambda} / m_\rho$ $g_{\omega n} / m_{\omega}$ and 
$g_{\omega \Lambda} / m_\omega$. 
The nucleon couplings are determined from the nuclear matter saturation 
properties \cite{3d,3e}. The values are given as 
follows: $C_\sigma^2 = 357.4$ and $C_\omega^2 = 273.8$ for the Walecka model; 
$169.2$ and $59.1$ for the ZM model; $219.3$ and $100.5$ for the ZM2 model 
and $443.3$ and $305.5$ for the ZM3 model where $C_\sigma^2 
= g^2_\sigma (M_N / m_\sigma)^2$ and $C_\omega^2 
= g_\omega^2(M_N / m_\omega)^2$. The value of $g_{\rho n}$ is 
$8.55451$. The other three parameters are coupling constants of the
hyperon-meson interactions and are not well
known. These cannot be determined from nuclear matter properties
since the nuclear matter does not contain hyperons. Furthermore,
properties of hypernuclei do not fix these parameters in
a unique way. Here the hyperon couplings are taken to be ${ 2 \over 3}$ of 
the nuclear coupling \cite{3c}.

The non-linear Walecka model has eight parameters out of which five 
are determined by the saturation properties of nuclear matter. These 
are nucleon couplings to scalar $g_\sigma / m_\sigma$, isovector 
$g_{\rho }/ m_\rho$ and vector mesons $g_\omega / m_\omega$ and  
the two coefficients in the scalar self interaction i.e. $b$ and $c$. 
Here the $\rho$ and $\omega$ meson masses are chosen to be their physical
masses.

The values of the parameters obtained are \\
\begin{center}
$g_{\sigma n} /m_\sigma = 1.52505 \times 10^{-2}$, $g_{\omega n} /m_\omega 
=  8.63433/782 $, $g_\rho /m_\rho = 8.55451/770$, \\
$b = 3.41799 \times 10^{-3}$ and $c = 1.46 \times 10^{-2}$. 
\end{center}
Once again the hyperon couplings are taken to be $2 \over 3$ of the nuclear 
coupling. 

\section {Quark Model}
The Colour Dielectric Model is 
based on the idea of Nielson and Patkos \cite{3b}. 
In this model, one  generates the confinement of quarks and gluons 
dynamically through the interaction of these fields with a scalar field. 
In the present work, the chiral extension (CCD) of this model has been 
used to calculate the quark matter equation of state \cite{3c}. The 
Lagrangian density of CCD model is given by \cite{3c}
\begin{eqnarray}
{\cal {L}} = \bar\psi(x)\big \{ i\gamma^{\mu}\partial_{\mu}-
(m_{0}+m/\chi(x) U_{5}) + (1/2) g
\gamma_{\mu}\lambda_{a}A^{a}_{\mu}(x)\big \}\psi \nonumber \\
+f^{2}_{\pi}/4 Tr ( \partial_{\mu}U
\partial^{\mu}U^{\dagger}) - 
1/2m^{2}_{\phi} \phi^{2}(x) -(1/4)
\chi^{4}(x)(F^{a}_{\mu\nu}(x))^{2} \nonumber \\ 
+(1/2)
\sigma^{2}_{v}(\partial_{\mu}\chi(x))^{2} -U(\chi)
\label{ccd}
\end{eqnarray}
\noindent where $U = e^{i\lambda_{a}\phi^{a}/f_\pi}$ and $U_{5} =
e^{i\lambda_{a}\phi^{a}\gamma_{5}/f_{\pi}}$, $\psi(x)$,
$A_{\mu}(x)$, $\chi(x)$ and $\phi(x)$ are quark, gluon, scalar (
colour dielectric ) and meson fields respectively, $ m_{\phi}$
and m are the meson and quark masses, $ f_{\pi}$ is the pion
decay constant, $F_{\mu\nu}(x)$ is the usual colour
electromagnetic field tensor, g is the colour coupling constant
and $\lambda_{a}$ are the Gell-Mann matrices.  The flavour
symmetry breaking is incorporated in the Lagrangian through the
quark mass term $(m_{0}+m/\chi U_{5})$, with $m_{0} = 0$ for $u$
and $d$ quarks. So the masses of $u$, $d$ and $s$ quarks are
$m$, $m$ and $m_{0} + m$ respectively. The self interaction 
$U(\chi)$ of the scalar field is assumed to be of the form
\begin{eqnarray}
U(\chi)= \alpha B
\chi^2(x)[
1-2(1-2/\alpha)\chi(x)+(1-3/\alpha)\chi^2(x)],
\end{eqnarray}
\noindent so that $U(\chi)$ has an absolute minimum at $\chi =
0$ and a secondary 
minimum at $\chi = 1$. Thus the parameter B is the analog of the 
bag pressure of the bag model. The interaction of the scalar 
field with quark 
and gluon fields is such that quarks and gluons can not exist in the
region where $\chi= 0$.
In  the  limit  of  vanishing meson masses, the 
Lagrangian of eqn.(\ref{ccd}) is invariant under chiral transformations 
of quark and meson fields. 

The parameters of the CCD model are the quark masses $m_q$ ($q = u, d, s$), 
strong coupling 
constant $\alpha_s$, the Bag pressure $B$ and the constant $\alpha$. These 
are obtained by fitting the baryon masses. In the present calculation 
we have used $m_{q(u,d)} = 92 MeV$, $m_{q(s)} = 295 MeV$, 
$B^{1/4} = 152 MeV$, $\alpha =36$ and $\alpha_s = 0.08$. 

The quark matter calculation proceeds as follows. It is assumed that in the 
presence of non-zero quark/anti-quark densities, the square of the meson 
field ($\langle \phi^2 \rangle$) may develop a non-zero vacuum expectation 
value. This assumption is analogous to the assumption that, in the 
linear-sigma model \cite{bron} the $\sigma$-field acquires non-zero vacuum 
expectation values. It has been found that this occurs when the quark 
density exceeds a certain critical value. As a result of this the effective 
quark masses decrease with the increase in $\langle \phi^2 \rangle$ and 
hence with increase in baryon density. Thus at large quark densities 
one obtains from CCD model an equation of state similar to the equation 
of state of free quarks and gluons. Furthermore, the colour dielectric field 
$\chi$ and $\langle \phi^2 \rangle$ in quark matter are evaluated using mean 
field approximation. Quark-gluon interaction is incorporated upto two loop 
in thermodynamic potential. 

\section {Hot and Dense Baryonic Matter}
The thermodynamic equilibrium in the hot and dense matter can be imposed 
from the weak 
interaction conditions. This introduces the following relations among 
different bare chemical potentials as,
\begin{eqnarray}
\mu_p &=& \mu_n - \mu_e \nonumber\\
\mu_\Lambda &=& \mu_n \nonumber\\
\mu_{\Sigma^-} &=& \mu_n + \mu_e \nonumber\\
\mu_\mu &=& \mu_e 
\end{eqnarray}

In the medium the baryon chemical potentials are modified as,
\begin{eqnarray}
{\bar \mu}_n &=& \mu_n - g_{\omega n} \omega_0  + {1 \over 2} g_\rho \rho_0 
\nonumber\\
{\bar \mu}_p &=& \mu_p - g_{\omega p} \omega_0  - {1 \over 2} g_\rho \rho_0 
\nonumber\\
{\bar \mu}_\Lambda &=& \mu_\Lambda - g_{\omega \Lambda} \omega_0 \nonumber\\
{\bar \mu}_{\Sigma^-} &=& \mu_{\Sigma^-} - g_{\omega \Sigma^-} \omega_0  
+ g_\rho \rho_0 
\end{eqnarray}
where ${\bar \mu}_i$ is the chemical potential of the $i$'th baryon in the 
medium and $\mu_i$ is the bare chemical potential for the same. 

The effective mass of a baryon in matter is as given in 
equations (11) and (12).

The charge neutrality gives,
\begin{eqnarray}
n_{p}= n_{e} + n_{\mu} + n_{\Sigma^-}
\end{eqnarray}

In addition the baryon number conservation gives, 
\begin{equation}
n_B = n_n + n_p + n_{\Lambda} + n_{\Sigma^-} 
\end{equation}

In the last two equations $n_i$ is the number density of the i'th particle 
and is given by 
\begin{equation} 
n_i = 2 \int {{d^3p} \over {(2\pi)^3}} \left [f_i^+ - f_i^-\right]
\end{equation}

According to the usual convention the baryon chemical potential is defined 
as $\mu_B = \mu_n$ \cite{1af}. 

All the above equations need to be solved self consistently to obtain 
the density dependence of baryon masses and number densities at 
different temperatures. In figures 1-3 the temperature dependence of 
$n$, $\Lambda$ and $\Sigma^-$ masses have been plotted for five different 
models which have been considered here and at three different densities - 
$\rho = 0, 0.15$ and $0.3 fm^{-3}$. The density dependence of baryon masses 
at three temperatures - $T = 0, 100$ and $200 MeV$ have been plotted in 
figures 4-6. The number densities for different models at 
$T = 0$ and $200 MeV$ have been plotted in figures 7-11. 

\section {Quark Hadron Phase Transition}
The quark-hadron phase transition can now be calculated by using the 
effective models discussed in section 2 and section 3. 
The phase transition point is determined by adopting Gibb's criterion. This 
says that the point at which the free energies ( or pressure ) of the two 
phases, for a given chemical potential, are equal corresponds to the phase 
transition point. In the present context, the baryon chemical potential 
($\mu_B$) is the only independent chemical potential. 
Then the crossing of the pressure curves for two phases in 
$P$ -$\mu_B$ plane gives the phase transition point. Glendenning, on the 
other hand, considered a case where, in the mixed phase, neutron and quark
matter are not charge neutral but the mixture as a whole is \cite{3f}. This 
aspect has been studied further by Heiselberg et.al. in ref \cite{3g}. This 
situation is not considered here. 

For a first order transition, the derivatives of the $P$- $\mu_B$
curve for the two phases at the phase transition point are not equal and the 
difference in the two derivatives gives the discontinuity in the density of 
the two phases at the transition point. The two phases coexist in this range 
of density. The latent heat of transition is given by the difference in 
the energy densities of the two phases at the critical point. As mentioned
earlier, the lattice QCD calculations indicate that hadron - quark phase 
transition may be weakly first order or second order. In a calculation, such 
as presented here, one would necessarily get a first order transition since 
two different models are employed to calculate the properties of quark and 
hadron phases. 

The pressure for Walecka and ZM models can be written as 
\begin{equation}
P = m^{* \alpha} {1 \over {2}} m_{\omega}^{2} \bar \omega_{0}^{2} 
+ m^{* \alpha} {1 \over{2}} m_{\rho}^{2}(\bar \rho_{0}^{3})^2 - 
{1 \over {2}} m_{\sigma_0}^{2} \bar \sigma^{2} 
+ \sum_{i} P_{FG}(\bar m_{i}, \bar \mu_{i}) +\sum_{l} P_{FG}(m_{l},
\mu_{l}) 
\end{equation}

For the Non-Linear Walecka model pressure $P$ is given by 
\begin{eqnarray}
P~={1 \over {2}} m_{\omega}^{2} \bar \omega_{0}^{2} + {1 \over
{2}} m_{\rho}^{2}(\bar \rho_{0}^{3})^2 - {1 \over {2}}
m_{\sigma_0}^{2} \bar \sigma_0^2 - {1 \over 3} bm_{N}(g_{\sigma 
N} \bar \sigma_0)^3 - {1 \over {4}}c(g_{\sigma N} \bar \sigma)^4
\nonumber \\
+ \sum_{i} P_{FG}(\bar m_{i}, \bar \mu_{i}) +\sum_{l} P_{FG}(m_{l},
\mu_{l}) 
\end{eqnarray}

In the above two equations $P_{FG}$ is the pressure for the free fermi gas 
and is given by
\begin{equation}
P_{FG} = {1 \over {3\pi^2}} \int p^4dp \sum_i {1 \over {E_i}} \left[f_i^+ + 
f_i^- \right]
\end{equation}

For the quark sector the pressure is calculated upto two loop in the 
quark-gluon interaction and is given by 
\begin{eqnarray}
P &=& T \gamma_q \int {{d^3p} \over {(2\pi)^3}} 
\left[ln(1+e^{(\mu-\epsilon_q(p))/T}) 
+ ln(1+e^{-(\mu+\epsilon_q(p))/T}) \right] \nonumber\\
&-& T \gamma_g \chi^4 \int {{d^3p} \over {(2\pi)^3}} 
ln(1+e^{-\epsilon_g(p)/T}) - T \sum_\phi \gamma_\phi \int {{d^3p} 
\over {(2\pi)^3}} 
ln(1-e^{-\epsilon_\phi(p)/T}) \nonumber\\
&-&{16 \over 3} \pi \alpha_s T^2 \int {{d^3p n(p)} \over {(2\pi)^3 \epsilon_q(p)}}
-8\pi \alpha_s \int {{d^3p d^3q} \over {(2\pi)^6 \epsilon_q(p) \epsilon_q(q)}} \nonumber\\
&\times& \{ \left[{{2m^{*2}} \over {(\epsilon_q(p) - \epsilon_q(q))^2 
- \epsilon_g^2}} -1 \right] \times \left[n^-(p) n^-(q) + n^+(p) n^+(q)\right]  
\nonumber\\
&+& \left[{{2m^{*2}} \over {(\epsilon_q(p) + \epsilon_q(q))^2 
- \epsilon_g^2}} -1 \right] 
\times \left[n^-(p) n^-(q) + n^+(p) n^+(q)\right] \} \nonumber\\
&-& {2 \over 3} \alpha_s \pi T^4 
\end{eqnarray}

In the above expression $\epsilon_q$, $\epsilon_g$ and $\epsilon_\phi$ are 
the quark gluon and meson kinetic energies, $\gamma_q$, $\gamma_g$ and 
$\gamma_\phi$ are the corresponding degeneracies, 
$n^-(p) = 1/(1+exp((\epsilon_q(p) - \mu)/T))$ and $n^+(p) 
= 1/(1+exp((\epsilon_q(p) + \mu)/T))$ are the quark and 
antiquark distribution functions respectively, $N(p) = 
1/(exp(\epsilon_\phi(p)/T) - 1)$ is the distribution function for mesons and 
$n(p) = n^+(p) + n^-(p)$. In the quark sector the chemical equilibrium 
under weak decay and the charge neutrality give 
\begin{equation}
\mu_d = \mu_u + \mu_e; {\hskip 0.1in} \mu_s = \mu_u + \mu_e
\end{equation}
and 
\begin{equation}
{2 \over 3} n_u - {1 \over 3} n_d - {1 \over 3} n_s - n_e = 0
\end{equation}
The baryon density $n_B = {1 \over 3} \sum_i n_i$ where $i = u, d, s$ and 
the baryon chemical potential is defined as $\mu_B = \mu_u + \mu_d + \mu_s$. 

The $P-\mu_B$ curves for different models have been plotted in figures 
12-16.

\section{Results and Discussions}
In the present work we have studied the properties of hot and dense 
asymmetric baryonic matter in $\beta$-equilibrium. The baryonic matter 
has been discussed within the framework of five effective hadronic models. 
They are the LW model \cite{1h}, three variants of the ZM model 
\cite{1ah} - ZM, 
ZM2 and ZM3 and the NW model \cite{1af}. We have extended the LW, ZM, ZM2 
and ZM3 models to include hyperons and also discussed the NW model. The 
models include four baryons - n, p, $\Lambda$ and $\Sigma^-$, the scalar 
$\sigma$-meson, the vector mesons $\omega$ and $\rho$ and the leptons 
(electron and muon). The baryonic matter properties {\it i.e.} the masses 
and fractional number densities have been calculated from these models in 
the mean field level. The pressure of the baryonic matter has been 
calculated in the usual canonical fashion.  
The pressure for quark matter has been calculated from CCD model \cite{3c}. 

The temperature dependence of baryon masses at different densities have 
been plotted in figures 1-3 for different models. Fig.1 is the temperature 
dependence of nucleon mass at different densities. This shows that the 
nucleon mass decreases with temperature and the maximum change in the 
nucleon mass is obtained in the LW model. Also, as the density is increased  
the baryon mass first increases slightly with temperature and then starts 
decreasing. The same qualitative behaviour is observed in the case of 
$\Lambda$ and $\Sigma^-$ masses (fig.2 and 3). With density, the variation 
of nucleon mass is even faster. In the LW model the nucleon mass goes to 
zero at about $0.8 fm^{-3}$ whereas in the other models the nucleon mass 
saturates in the range $400 - 600 MeV$. The higher value is for ZM model 
and lower is for ZM3 model. 

In figures 7-11 the variation of fractional number densities,  
($n_i/n_B$), of different particles with baryon density for different 
models have been plotted, each at two temperatures $T = 0 MeV$ and 
$T = 200 MeV$. Fig. 7 shows that the neutron number density in the LW model 
decreases and goes to zero at about $0.8 fm^{-3}$. The proton number density 
increases initially from $0$ to $0.25 fm^{-3}$ then it decreases slightly 
upto $1.0 fm^{-3}$ after which it saturates. The $\Lambda$ baryon appears at 
about $n_B =  0.25fm^{-3}$, increases till $1.0 fm^{-3}$ to about $0.95$ and 
then saturates. The $\Sigma^-$ also appears at $n_B = 0.25 fm^{-3}$, it 
increases first and then starts decreasing going to zero at 
$n_B = 1.1 fm^{-3}$. The matter becomes hyperonic at $n_B =  0.4 fm^{-3}$. 
(We will use this terminology of "hyperonic matter", in the discussion of 
number densities, for the situation when the $\Lambda$ or $\Sigma^-$ number 
density takes over $n$ or $p$ number density). In the same model, at 
$T = 200 MeV$ the matter is hyperonic right from the begining. The $\Lambda$ 
number density varies from $0.39$ to $0.85$ as the density is varied from 
$0$ to $1.5 fm^{-3}$. 

The nature of variation of number densities is somewhat different for 
the ZM model (fig. 8). At 
$T = 0 MeV$, the neutron number density decreases with baryon density but 
does not go to zero. The proton number density saturates at a higher 
value compared to that in the LW model. The transition from nuclear to 
hyperonic matter takes place at $n_B = 1.1 fm^{-3}$ which is considerably 
higher than that in the LW model. In the ZM model, at $T = 200 MeV$, 
unlike the LW model, the matter is not hyperonic at the very begining. The 
transition takes place at about $n_B = 1.6 fm^{-3}$. For ZM2 model 
(fig. 9), at $T = 0 MeV$, the transition takes place at $n_B = 
1.6 fm^{-3}$ whereas at $T = 200 MeV$ it is at $n_B = 1.9 fm^{-3}$. 
Figure~10 gives an interesting result. This is for the ZM3 model. The 
matter does not become hyperonic ever. It stays as nuclear matter at all 
temperatures and densities. In the NW model at $T = 0 MeV$ the transition 
occurs at $n_B = 0.85 fm^{-3}$. Unlike LW model the neutron and sigma 
number densities do not go to zero. The proton number density also 
saturates at a higher value compared to that in the LW model. 

The hadron-quark phase transition has been studied using the Gibbs' 
criterion. The intersection of the two $P-\mu_B$ curves, for two different 
models for the two sectors, gives the transition density. We have a first 
order phase transition due to the consideration of two different models in 
two phases. In figures 12-16 we have plotted the $P-\mu_B$ curves for five 
different hadronic models along with the quark model at five different 
temperatures. The quark-hadron phase transition, for this set of models, is 
observed only 
for LW and NW models and that too at $T = 0 MeV$ and $T = 50 MeV$. The 
transition, for the LW model at $T = 0 MeV$, takes place at $\mu_B = 1100 
MeV$ which corresponds to $\rho_c = 0.27 fm^{-3}$. At $T = 50 MeV$ the critical 
density is $\rho_c = 0.26 fm^{-3}$. For the NW model these two values are  
$\rho_c = 0.47 fm^{-3}$ and $\rho_c = 0.37 fm^{-3}$
respectively. 

To conclude, we have compared different hadronic models at finite temperature 
and density. These models give varying predictions for masses and number 
densities. This will have a strong bearing both for neutron star properties 
as well as heavy ion collisions in the laboratory. Also the neutrino 
emissivities will be different for these models. 

The hadron-quark phase transition has been investigated using CCD model. The 
phase transition is seen for LW and NW whereas none of the variants of ZM 
model gives the phase transition for the parameter set considered here, 
even at $T = 0 MeV$. This will imply the absence of quark-hadron phase 
transition inside neutron star. 

\vskip 0.1in
\noindent
{\bf Acknowledgement} 

AB would like to thank Department of Atomic Energy (Govt. of India) and
SKG would like to thank Council of Scientific and Industrial Research 
for financial support.


\newpage
\begin{thebibliography}{99} 
\bibitem{1a} A very good review on hot and dense hadronic matter may  
be obtained in : B.D.Serot and J.D.Walecka, Adv. Nucl. Phys. {\bf 16}, 1 
(1986) and references therein.
\bibitem{1j} J.Collins and M.Perry, Phys.Rev.Lett. {\bf 34}, 135 (1975). 
\bibitem{1k} E.V.Shuryak, Phys.Rep. {\bf 61}, 71 (1980). 
\bibitem{1h} J.D.Walecka, Ann.Phys. {\bf 83}, 491 (1974).
\bibitem{3a} S.A.Chin, Ann.Phys. (N.Y.), {\bf 108}, 301 (1977).   
\bibitem{1af} J.I.Kapusta and K.A.Olive, Phy.Rev.Lett. {\bf 64}, 13 (1990); 
J.Ellis, J.I.Kapusta and K.A.Olive, Nucl.Phys. {\bf B348}, 345 (1991);
\bibitem{1ah} J.Zimanyi and S.A.Moszkowski, Phys. Rev. {\bf C42}, 1416 
(1990).   
\bibitem{3d} A.Delfino, C.T.Coelho and M.Mallherio,Phys.Lett. {\bf B345}, 
361 (1995). 
\bibitem{1ag} S.K.Ghosh, S.C.Phatak and P.K.Sahu, Z.Phys. {\bf A352}, 457 
(1995). 
\bibitem{1ai} N.K.Glendenning, F.Weber and S.A.Moszkowski, Phys.Rev. 
{\bf C45}, 844 (1992). 
\bibitem{1aj} V.Thorsson, M.Prakash and J.M.Lattimer, Nucl.Phys. {\bf A572}, 
693 (1994). 
\bibitem{3e} A.Bhattacharyya and S.Raha, Phys.Rev. {\bf C53}, 522 (1996).  
\bibitem{3c} S.K.Ghosh, "Study of Quark Matter in Chiral Colour Dielectric 
Model and Its Application to Dense Stars", Ph.D. Thesis, and references 
therein. 
\bibitem{3b} H.B.Nielson and A.Pattkos, Nucl.Phys. {\bf B195}, 137 (1982). 
\bibitem{bron} W.Broniowski, M.Cibej, M.Kutschera and M.Rossina, Phys. Rev. 
{\bf D41}, 285 (1990).  
\bibitem{3f} N.K.Glendenning, F.Weber and S.A.Moszkowski, Phys.Rev. 
{\bf C45}, 844 (1992).
\bibitem{3g} H.Heiselberg, C.J.Pethick and E.F.Staubo, Phys.Rev.Lett. 
{\bf 70}, 1355 (1993).
\end{thebibliography}
\end{document}